\documentstyle[10pt, amscd]{amsart}
\newtheorem{theo}{Theorem}
\newtheorem{prop}[theo]{Proposition}

\newtheorem{cor}[theo]{Corollary}

\newcommand{\Rr}{{\rm I\!R}}

\newcommand{\half}{\frac{1}{2}}

\newcommand{\C}{{\mathchoice {\setbox0=\hbox{$\displaystyle\rm C$}\hbox{\hbox
to0pt{\kern0.4\wd0\vrule  height0.9\ht0\hss}\box0}}
{\setbox0=\hbox{$\textstyle\rm C$}\hbox{\hbox
to0pt{\kern0.4\wd0\vrule  height0.9\ht0\hss}\box0}}
{\setbox0=\hbox{$\scriptstyle\rm C$}\hbox{\hbox
to0pt{\kern0.4\wd0\vrule  height0.9\ht0\hss}\box0}}
{\setbox0=\hbox{$\scriptscriptstyle\rm C$}\hbox{\hbox
to0pt{\kern0.4\wd0\vrule  height0.9\ht0\hss}\box0}}}} 

\newcommand{\Z}{{{\mathchoice  {\hbox{$\textstyle  Z\kern-0.4em  Z$}}
{\hbox{$\textstyle  Z\kern-0.4em  Z$}}
{\hbox{$\scriptstyle  Z\kern-0.3em  Z$}}
{\hbox{$\scriptscriptstyle  Z\kern-0.2em  Z$}}}}}

\begin{document}
\title[Szego kernels and a  theorem of Tian]
{Szego kernels and a  theorem of Tian}
\thanks{Partially supported by  NSF grant
\#DMS-9703775}
\author{Steve Zelditch}
\address{Department of Mathematics, The Johns Hopkins University, Baltimore,
Maryland 21218}

\maketitle

\section{Introduction}  A variety of results in complex geometry and mathematical
physics depend upon the analysis of holomorphic sections of high powers $L^{\otimes
N}$ of holomorphic line bundles $L \rightarrow M$ over compact K\"ahler manifolds
(\cite{A}\cite{Bis}\cite{Bis.V} \cite{Bou.1}\cite{Bou.2}\cite{B.G}\cite{D}\cite{Don} \cite{G}\cite{G.S}\cite{K}\cite{Ji} \cite{T} \cite{W}). The principal tools have
been H\"ormander's $L^2$-estimate on the $\bar{\partial}$-operator over $M$ [T], the
asymptotics of heat kernels $k_N(t,x,y)$ for associated Laplacians
\cite{Bis}\cite{Bis.V}\cite{D}\cite{Bou.1}\cite{Bou.2}\cite{G}, the method of stationary phase for formal functional
integrals \cite{A}\cite{W} and the microlocal analysis of Szeg\"o and Bergman kernels
\cite{B.F.G}\cite{B.S}\cite{B.G}.  In this note we wish to apply the latter methods, specifically
the Boutet de Monvel-Sj\"ostrand parametrix for the Szeg\"o kernel, to
a problem in complex geometry.  Our purpose is to prove the following theorem: 

\begin{theo}\label{MAIN} Let $M$ be a compact complex  manifold of dimension $n$ (over $\C$) and
let $(L, h) \rightarrow M$ be a positive hermitian holomorphic line bundle. Let
$g$ be the K\"ahler metric on $M$ corresponding to the K\"ahler form $\omega_g:=
Ric(h)$. For each $N \in {\bf N}$, $h$ induces a hermitian metric $h_N$ on
$L^{\otimes N}$. Let $\{S_0^N, \dots, S_{d_N}^N \}$ be any orthonormal basis of
$H^0(M, L^{\otimes N})$, $d_N = \rm{dim} H^0(M, L^{\otimes N})$,  with respect to the
inner product  $\langle s_1, s_2 \rangle_{h_N} = \int_M h_N(s_1(z), s_2(z)) dV_g$.
Here, $dV_g = \frac{1}{n!} \omega_g^n$ is the volume form of $g$.
Then there exists a complete asymptotic expansion:
$$ \sum_{i=0}^{d_N} ||S^N_i(z)||_{h_N}^2 =  a_0 N^n +
a_1(z) N^{n-1} + a_2(z) N^{n-2} + \dots$$
for certain smooth coefficients $a_j(z)$ with $a_0 = 1$.
 More precisely, for any $k$ 
$$|| \sum_{i=0}^{d_N} ||S^N_i(z)||_{h_N}^2 - \sum_{j < R} a_j(x) N^{n-j}||_{C^k}
\leq C_{R,k} N^{n- R}.  $$\end{theo}

Above, $Ric (h)$ is the Ricci curvature of $h$, given locally by    $\frac{\sqrt{-1}}{2\pi} \partial \bar{\partial} a$ where $a = ||e_L||_h$ is the positive function locally
representing $h$ in a local holomorphic frame $e_L$.  

This theorem has a number of corollaries.
First it implies that for sufficiently large $N$, there
are no common zeroes of the sections $\{S_0^N, \dots, S_{d_N}^N \}$. Hence
one can define the holomorphic map
 \begin{equation}\phi_N: M \rightarrow \C {\bf P}^{d_N},\;\;\;\;\; z \mapsto
[S_0^N(z), \dots, S_{d_N}^N(z)]\end{equation}
where $[S_0^N(z), \dots,
S_{d_N}^N(z)]$ denotes the line thru $(S_0^N(z), \dots,
S_{d_N}^N(z))$ as defined in a local holomorphic frame. Since all the components
transform by the same scalar under a change of frame, the line is well-defined.
As is well-known \cite{G.H}, $\phi_N$ is  equivalent to the invariantly
defined map
\begin{equation} \tilde{\phi}_N : M  \rightarrow {\bf P} H^0(M, L^{\otimes
N})^*,\;\;\;\;\; z \mapsto H_z := \{s \in  H^0(M, L^{\otimes N}) : s(z) =
0\}.\end{equation}

Secondly it gives an asymptotic formula for the distortion function between
the metrics $h_N$ and $h_{FS,N}$,  where $h_{FS, N}$ is the Fubini-Study metric
on $L^{\otimes N}$ induced by $\tilde{\phi}_N$. The following result was simeltaneously
 proved in the case of abelian varieties   by G.Kempf \cite{K} and S.Ji \cite{Ji} (with
$C^0$ convergence)  and
for general projective varieties  (with $C^4$ convergence) by    G.Tian
(\cite{T}, Lemma 3.2(i)). Heat
kernel proofs were later  found by T. Bouche \cite{Bou.1}\cite{Bou.2} and J.P.Demailly 
\cite{D}. 

\begin{cor}\label{DEMAILLY} Let $G$ be any Riemannian
metric on $M$, endow $H^0(M, L^{\otimes N})$ with the Hermitian inner product
induced by $(G, h_N)$ and define the map $\tilde{\phi}_N$ as above.
  Identify $L^{\otimes N}$ with the
pull-back $\tilde{\phi}_N^* O(1)$ of the hyperplane bundle $O(1) \rightarrow {\bf P}
H^0(M, L^{\otimes N})$, and  let $h_{FS, N}$ be the pullback of the standard
Hermitian metric on $O(1).$ 
   Then
$$ \frac{h_{N, z}}{h_{FS, N, z}} =  (\frac{N}{2\pi})^n
|\alpha_1(z) \dots \alpha_n(z)| + O(N^{n-1}).$$
where $\alpha_1(z), \dots, \alpha_n(z)$ are the eigenvalues $Ric(h)$
with respect to $G$. \end{cor}

 If we take the background metric $G$ to be the K\"ahler
metric associated to the K\"ahler form
$\omega_g = Ric (h)$, then the curvature eigenvalues are all equal to one.

Third, it implies:    
\begin{cor}\label{TIAN} Let $\omega_{FS}$ denote the Fubini-Study form on $\C {\bf P}^{d_N}$.
Then: $$||\frac{1}{N}  \phi_N^*(\omega_{FS}) - 
\omega_g||_{C^k} = O(\frac{1}{N})$$ for any $k$.  \end{cor}

This statement for $k \leq 2$ was the principal result of Tian (\cite{T}, Theorem A).
The map $\phi_N$ depends on the choice of $\{S_0^N, \dots, S_{d_N}^N \}$ but
it is easily seen that $ \phi_N^*(\omega_{FS})$ does not. This result follows
formally from the preceding one by taking the curvature of both sides of 
the asymptotic formula.

Our result strengthens the previous ones in two ways: First, it shows that
the convergence of $\frac{1}{N}  \phi_N^*(\omega_{FS}) \rightarrow \omega_g$
takes place in the $C^{\infty}$-topology and not just in $C^2$. This was
conjectured in \cite{T}. Second, it
shows that this convergence is just the first term of a complete asymptotic expansion.
  If only the principal terms are
desired, then the proof could be simplified further: as in \cite{Z}, which treats an
 analogous problem in the context of Zoll manifolds, one could obtain
the principal terms by the symbol calculus of Toeplitz operators.
But we believe that the lower order terms should also be  of interest.

The proof  begins by expressing the maps $\phi_N$ and $\tilde{\phi}_N$ in terms of an
associated equivariant map $\Phi_N$ on the unit circle bundle $X$ of the dual
line bundle $L^*$ with
respect to the induced metric $h$.  Roughly, this converts the holomorphic geometry of
$L$ to the CR geometry of $X$.  Since $X$ is the boundary of the strictly
pseudo-convex domain
 $D = \{v \in L^*: |v|_h < 1\} \subset L^*$ it has a Szeg\"o kernel $\Pi(x,y)$ which
projects $L^2(X)$ to the Hardy space $H^2(X)$ of boundary values of holomorphic
functions in $D$.  Under the natural $S^1$ action of $X \rightarrow M$, $H^2(X)$
splits up into weight spaces $H^2_N(X)$ and one has a canonical isomorphism
$s \rightarrow \hat{s}:  H^0(M, L^{\otimes N})\mapsto H^2_N(X) $,  $\hat{s}(z, u) =
\langle u, s(z) \rangle$ where we write a point of $X$ as $(z,u), u \in L_z^*, |u|_h
= 1$ and where $\langle \cdot, \cdot \rangle$ is the pairing between $L$ and $L^*.$ 
 So the basis $\{S_0^N, \dots,
S_{d_N}^N \}$ of $H^0(M, L^{\otimes N})$ corresponds to an orthonormal basis
$\{\hat{S}_0^N, \dots,
\hat{S}_{d_N}^N \}$
of $H^2_N(X)$. One then gets  an associated CR map of $X$ into
$H^2_N(X)^*$.  Expressed invariantly in terms of the orthogonal projection
$\Pi_N$ onto $H^2_N(X)$ it is defined by:  
 \begin{equation} \Phi_N(x) = \Pi_N(x,
\cdot): X \rightarrow H^2_N(X)^*. \end{equation}
Using the canonical isomorphism above we  get an essentially equivalent map
$\tilde{\Phi}_N : X \rightarrow H^0(M, L^{\otimes N})^*$.  (We note that these maps
are well-defined even when the set $Z_N$ of common zeroes of the sections is
non-empty.) Thus for $N \gg 0$ we get the diagram:
\begin{equation} \begin{array}{lll} X & \stackrel{\tilde{\Phi}_N }{\rightarrow} & H^0(M,
L^{\otimes N})^* - 0 \\ \pi \downarrow & & \downarrow \rho \\ M & 
\stackrel{\tilde{\phi}_N}{  \rightarrow} & 
{\bf P} H^0(M, L^{\otimes N})^* \end{array} \end{equation}
 where $\pi$ and $\rho$ are the canonical projections. The top arrow is well-defined
for all $N$.

After unravelling the identifications, we find (\S 1) that
$\sum_{i=0}^{d_N} ||S^N_i(z)||_{h_N}^2 = \Pi_N(x,x)$  and that 
   $\frac{1}{N} \phi_N^* \omega_{FS} = \omega_g +  \frac{i}{2\pi} \bar{\partial}_b
\partial_b \log \Pi_N (x,x)$ for any $x$ with $\pi(x) = z$. The second term on
the right side is an $S^1$ invariant form so we have identified it with a form on
$M$. The 
theorem is therefore equivalent to the statement that $\Pi_N(x,x)$ has a complete
asymptotic expansion as $N \rightarrow \infty$ which can be differentiated any number
of times.  This will follow by applying the method of stationary phase to
 Boutet de Monvel-.Sj\"ostrand's  parametrix for the  Szeg\"o projector $\Pi(x,y)$\cite{B.S}
(see \S 3).

The analysis of Szeg\"o kernels should have other applications in complex geometry.
By studying the off-diagonal of $\Pi_N(x,y)$ one can show that the maps
$\Phi_N$ are embeddings, thus obtaining an analytic proof of the Kodaira embedding
theorem.  In a forthcoming paper \cite{S.Z}, B.Shiffman and the author also use the
Szeg\"o kernels to show (among other things) that
the zeroes of a `random section' of $H^0(M, L^{\otimes N})$ become uniformly
distributed as $N \rightarrow \infty.$  There are also some potential analogues
in the almost complex setting. In  a recent paper \cite{B.U},  Borthwick-Uribe conjecture
some results on Szeg\"o kernels in the almost complex setting which seem very close
to what is proved here in the complex setting. In part their motivation ( as well
as ours) was to reinterpret some constructions of
Donaldson \cite{Don} from the viewpoint of semiclassical analysis.
 
We thank  B.Shiffman for  help with  the relevant complex geometry
and for his and M. Zworski's encouragement to publish this note.

\section{From line bundle to circle bundle}

The purpose of this section is to convert the statements of Theorem \ref{MAIN} and
Corollaries \ref{DEMAILLY} - \ref{TIAN} into statements about $\Pi_N.$
Before doing so let us recall why one exists in this context and establish some
notation.

\subsection{The CR setting}

 Let $O(1) \rightarrow \C {\bf P}^n$ denote the hyperplane section line bundle and let
$\langle \cdot,  \cdot \rangle$ denote its natural Hermitian metric. Let $M \subset
\C P^n$ be a non-singular projective variety, let $L$  denote the restriction
of $O(1)$ to $M$ and let $h$ denote the restriction of $\langle
\cdot,  \cdot \rangle$ to $L$.  The following proposition is well-known (it was
originally observed by Grauert in the 50's):

\begin{prop} Let $D = \{(m, v) \in L^* : h(v,v) \leq 1\}$. Then $D$ is a strictly
pseudoconvex domain in $L$.\end{prop}

Here $L^*$ is the dual line bundle to $L$. 
The boundary of $D$ is a principal $S^1$ bundle $X \rightarrow M$ whose defining
function is given by 
\begin{equation} \rho: L^* \rightarrow \Rr,\;\;\;\;\;  \rho(z,\nu) =
1 - |\nu|_z^2 \end{equation}
 where $\nu \in
L_z^*$ and where  $|\nu|_z$ is its norm in the metric induced by $h$. That is,
$D = \{ \rho > 0\}$. In a local
coframe $e_L^*$ over $U \subset M$ we may write $\nu = \lambda e_L^*$ and then
$|\nu|^2_z = a(z) |\lambda|^2$ where $a(z) = |e_L^*|^2_z$ is a positive smooth
function on $U$. Thus in  local holomorphic coordinates $(z, \lambda)$ on $L^*$ the
defining function is given by $\rho = 1 - a(z) |\lambda|^2.$   
We will denote the $S^1$ action by $r_{\theta} x$ and its infinitesimal generator 
by $\frac{\partial}{\partial \theta}$. We note that $\rho$ is $S^1$-invariant.

Let us denote by $T'D, T'' D \subset T D \otimes \C$ the holomorphic, resp.
anti-holomorphic subspaces and define $d'f = df|_{T' }, d'' f = 
df|_{T''}$ for $f \in C^{\infty}(D).$ Then $X$ inherits a CR structure 
  $TX \otimes \C = T' \oplus T''
\oplus \C \frac{\partial}{\partial \theta}$. Here $T' X$ (resp. $T''X$) denotes the
holomorphic (resp. anti-holomorphic vectors) of $D$ which are tangent to $X$. They
are given in local coordinates by vector fields $\sum a_j \frac{\partial}{\partial
\bar{z}_j}$ such that $\sum a_j \frac{\partial}{\partial \bar{z}_j} \rho = 0.$   A
local basis is given by the vector fields $Z_j^k = \frac{\partial}{\partial \bar{z}_j
} - (\frac{\partial \rho}{\partial \bar{z}_k })^{-1} (\frac{\partial \rho}{\partial
\bar{z}_j }) \frac{\partial}{\partial \bar{z}_k }$ ($j \not= k.$)

The Cauchy-Riemann operator on $X$ is defined by
\begin{equation} \bar{\partial}_b : C^{\infty}(X) \rightarrow C^{\infty}(X, (T'')^*),
\;\;\;\;\;\; \bar{\partial}_b  f = df |_{T''}. \end{equation}
In terms of the local basis above, it is given by 
\begin{equation} \bar{\partial}_b f = \sum_{j \not= k} Z_j^k f d\bar{z}_j |_{T''}.
\end{equation}

Also associated to $X$ are 
\begin{equation}\begin{array}{ll} \bullet & \rm{ the\;\; contact\;\; form}\;\;\alpha =
 \frac{1}{i} d'
\rho|_X = -  \frac{1}{i} d'' \rho|_X\\ &
\\ \bullet & \rm{ the\;\; volume\;\; form}\;\;d\mu = \alpha \wedge (d\alpha)^n \\
& \\ \bullet &
\rm{the \;\;Levi\;\;form}\;\; L_{\rho}(z)  = \sum \frac{\partial^2 \rho}{\partial
z_j \partial \bar{z}_k} z_j \bar{z}_k.\\
\\\bullet & 
\rm{the \;\;Levi \;\;form \;\;on}\;\; X\;\;L_X = L_{\rho}|_{T' \oplus T'' 
\cap TX}
 \end{array}  \end{equation} which are independent of the choice of $\rho.$
The Levi form on $X$ is related to $d \alpha = \pi^* \omega_g$ by:
$L_X(V,W) = d\alpha (V, \bar{W}).$  Since $\omega_g$ is
K\"ahler, $D$ is a strictly pseudoconvex domain.

The Hardy space $H^2(X)$ is the space of boundary values of holomorphic functions
on $D$ which are in $L^2(X)$, or equivalently $H^2 = (\rm{ker}\;\; \bar{\partial}_b)
\cap L^2(X).$   The $S^1$ action commutes with $\bar{\partial}_b$, hence $H^2(X) =
\oplus_{N = 1}^{\infty} H^2_N(X)$ where $H^2_N(X) = \{ f \in H^2(X): f(r_{\theta}x)
= e^{i N \theta} f(x) \}.$

A section $s$ of $L$ determines an equivariant function
$\hat{s}$ on $L^* - 0$ by the rule: $\hat{s}(z, \lambda) = \langle \lambda, s(z)
\rangle$ ( $z \in M, \lambda \in L^*_z.$) It is clear that if $\tau \in \C^*$
then $\hat{s}(z, \tau \lambda) =  \tau \hat{s}.$  We will usually restrict
$\hat{s}$ to $X$ and then the equivariance property takes the form:
$\hat{s}(r_{\theta} x) = e^{i  \theta}\hat{s}(x).$  Similarly, a section $s_N$
of $L^{\otimes N}$ determines an equivariant function $\hat{s}_N$ on $L^*-0$: put
$\hat{s}_N(z, \lambda) = \langle  \lambda^{\otimes N}, s_N(z) \rangle$ where
$\lambda^{\otimes N} = \lambda \otimes \lambda \otimes \dots \otimes \lambda.$
The following proposition is well-known:

\begin{prop} The map $s \mapsto \hat{s}$ is a unitary equivalence between
$H^0(M, L^{\otimes N})$ and $H^2_N(X).$ \end{prop}

As above, we let $\Pi_N : L^2(X) \rightarrow H^2_N(X)$ denote the orthogonal
projection.  Its kernel is defined by
\begin{equation} \Pi_N f(x) = \int_X \Pi_N(x,y) f(y) d\mu (y). \end{equation}
This definition differs from that of [B.S] in using $d\mu$ as the reference density.

\subsection{Line bundles and maps to projective space}

Since the definitions of the various maps $\phi_N, \tilde{\phi}_N,\Phi_N,
\tilde{\Phi}_N$ involve some identifications, we pause to recall some basic
facts about maps to projective space [\cite{G.H}, I.4]. 

Let $E \rightarrow M$ denote a  holomorphic line bundle. Since
we are interested in $E = L^{\otimes N}$ for large $N$ we may assume that not all
sections $s \in H^0(M,E)$ vanish at any point $z \in M.$ Then the space of sections
vanishing at $z$ forms a hyperplane $H_z$ in $H^0(M,E)$ and one can define a map
$\iota_E : M \rightarrow {\bf P}(H^0(M,E))^*$ by $z \rightarrow H_z.$
Here ${\bf P}(H^0(M,E))^*$ denotes the dual projective space of linear functionals
on $H^0(M,E)$ modulo scalar multiplication.

Now equip $E$ with a Hermitian metric $h$ and $M$ with a volume form, and
let $\langle, \rangle$ denote the induced inner product on $H^0(M,E)$.
Then choose an orthonormal basis $\{s_0, \dots, s_m\}$ of $H^0(M,E)$
with respect to $\langle, \rangle.$ Also, choose a
local holomorphic frame $e_E$ and write $s_j = f_j e_E.$ Then the point of
${\bf P}^m$ with homogeneous coordinates $[f_0(z) , \dots,
f_m(z) ]$ is independent of $e_E$ and defines a map $\phi_E : M \rightarrow \C {\bf
P}^m$.  The same basis also gives an identification
${\bf P}^m  \equiv {\bf P}(H^0(M,E))^* $ by writing a linear functional in the dual
basis $\{s_0^*, \dots, s_m^*\}$ of $H^0(M,E)^*.$ We observe that under this
identification, $\phi_E \equiv \iota_E:$ for   $H_z = \{\sum_j a_j s_j : \sum_j a_j
f_j(z) = 0\}  = \rm{ker}\;\;(\sum_j f_j(z) s_j^*) \leftrightarrow [f_0(z), \dots,
f_m(z)].$

Next, recall that the K\"ahler form $\omega_{FS}$ of the Fubini-Study metric
$g_{FS}$ on $\C P^{m}$ is given in homogeneous coordinates $[w_0, \dots, w_m]$
by $\omega_{FS} = \frac{\sqrt{-1}}{2\pi}   \partial \bar{\partial} \log
(\sum_{j=0}^{m} |w_i|^2).$  Hence
\begin{equation}  
\phi_E^*\omega_{FS} = \frac{\sqrt{-1}}{2\pi } \partial \bar{\partial}  \log
(\sum_{j=0}^{m}  |f_j|^2).  \end{equation}
It is easy to see that this form is independent of the choice of orthonormal
basis.

\subsection{The maps $\phi_N, \tilde{\phi}_N,\Phi_N, \tilde{\Phi}_N$}

Now let us return to our setting.  We fix a   local holomorphic section
 $e_L$ of $L$ over $U \subset M$. It induces
sections $e_L^N$ of $L^{\otimes N}|_U$ and  we  write $S_i^N (z) = f_i^N(z)
e^N_L(z)$ for a holomorphic function $f_i^N$ on $U$. By the above we have:
\begin{equation} \tilde{\phi}_N^* (\omega_{FS}) = \phi_N^*(\omega_{FS}) =    \frac{\sqrt{-1}}{2\pi } \bar{\partial}  \partial
\log (\sum_{j=0}^{d_N}  |f_j^N|^2).\end{equation}
 The definition is independent of the
choice of $e_L^N.$

  Since $S_i^N$ is a  holomorphic section of $L^{\otimes N}$, $\hat{S}_i^N$ is
an equivariant CR function of level $N$, i.e. $\hat{S}_i^N \in H^2_N(X).$
We will need a series of formulae relating expressions in $S_i^N$ to expressions
in $\hat{S}_i^N.$
Let us introduce local coordinates $(z, \theta)$ on $X$ on the domain $U$ of the
unitary frame $ \frac{e_L}{|e_L|}$ by $(z, \theta) \mapsto (z, r_{\theta}
\frac{e_L}{|e_L|}).$    

\begin{prop}  $||S_j^N(z)||_{h_N}^2 = |\hat{S}_i^N(x)|^2$ for any 
$x$ with $\pi (x) = z.$ \end{prop}

\noindent{\bf Proof}:

By definition,
 \begin{equation} \hat{S}_i^N (z, u) = \langle u^N , S_i^N(z)\rangle =
f_i^N(z) \langle u^N , e_L^N(z) \rangle = f_i^N(z) a^{\frac{N}{2}}(z) \langle u ,
\frac{e_L}{|e_L|}(z) \rangle^N. \end{equation}

In the above local coordinates, we get
$\hat{S}_i^N(z, \theta) = f_i^N (z) a(z)^{\frac{N}{2}} e^{i N \theta}.$ 
Hence $|\hat{S}_i^N(z, \theta)|^2 = a(z)^N |f_i^N (z)|^2.$ This obviously equals
$||S_j^N(z)||_{h_N}^2 $. \qed

\begin{prop} $\{\hat{S}_N^i\}$ is an orthonormal basis of $H^2_N(X).$ \end{prop}

\noindent{\bf Proof}:

Let $dV_g = \omega_g^n$ be the volume form of $(M,g)$. Then we have:
$$\begin{array}{l} \langle S_{\alpha}^N, S_{\beta}^N \rangle := \int_M h_N(
S_{\alpha}^N, S_{\beta}^N )d V_g = \int_M  a^N (z) f_i^N (z) \bar{f}_j^N (z) dV_g \\
\\
= \int_X  \hat{S}_{\alpha}^N \hat{S}_{\beta}^{*N } d
\mu\end{array}$$ where $d\mu = \alpha \wedge d\alpha^{n}$. In the latter step we use
that $\alpha \wedge d\alpha^{n} = d\theta \wedge \pi^* \omega_g^n.$ This follows from
the fact $d\alpha = \pi^* \omega$ and that $\alpha = d \theta + \eta$ where
$\eta$ only involves $dz, d\bar{z}.$ \qed

We further have:

\begin{prop}\label{BERGMAN} $\frac{1}{ N} \phi_N^* \omega_{FS}  = \omega_g +
\frac{\sqrt{-1}}{2\pi N} \partial \bar{\partial} \log (\sum_{j=0}^{d_N} 
||S_j^N||_{h_N}^2 )  = \frac{\sqrt{-1}}{2\pi N}\partial_b \bar{\partial}_b \log
(\sum_{j=0}^{d_N}  |\hat{S}_j^N|^2) + \omega_g.$ \end{prop}

\noindent{\bf Proof}: 

The first statement follows by writing $||S_j^N(z)||_{h_N}^2  = a^N(z)|f_j^N(z)|$ and
using that $\frac{\sqrt{-1}}{2\pi N }\partial \bar{\partial} \log a^N =  \omega_g.$ To prove the second
statement, we note that $\sum_{j=0}^{d_N}  |\hat{S}_j^N|^2$ and
$\partial_b \bar{\partial}_b \log
(\sum_{j=0}^{d_N}  |\hat{S}_j^N|^2)$
are $S^1$-invariant
and hence may be identified with  functions on $M$. 
In the latter case, this uses
the fact that the $S^1$ action commutes with $\partial_b,$ i.e. acts by CR 
automorphisms.
The statement then follows from the general fact that
$\pi_* \partial_b \pi^* f = \partial f$ for any $f \in C^{\infty}(M),$
 where $\pi_* F$ denotes
the function on $M$ corresponding to an $S^1$-invariant function $F$ on $X$. \qed

Now let us rewrite these relations in terms of the 
Szeg\"o projectors $\Pi_N$.

\begin{prop}\label{BERGMANSZEGO} $\frac{1}{ N} \phi_N^* \omega_{FS} = \omega_g + 
\frac{\sqrt{-1}}{2\pi N } \partial_b \bar{\partial}_b \log \Pi_N(x,x).$ \end{prop}

\noindent{\bf Proof}:

 We first observe that 
\begin{equation} \Pi_N(x,y) = \sum_{i=0}^{d_N} \hat{S}_i^N(x)\hat{S}^{N
*}(y) \end{equation}
or, in local coordinates, 
\begin{equation} \Pi_N(z, \theta, w, \theta') = a(z)^{\frac{N}{2}}
a(w)^{\frac{N}{2}}e^{i N (\theta -
\theta')} \sum_{i=0}^{d_N} f_i^N(z) \bar{f}_i^{N }(w) . \end{equation}
 Hence we have  \begin{equation} \begin{array}{ll} (a) & \sum_{j =
0}^{d_N} ||S_j^N(z)||_{h_N}^2 =  \Pi_N (z,0, z,0) \\ & \\(b) &
 \frac{\sqrt{-1}}{2\pi N} \partial \bar{\partial} \log (\sum_{j=0}^{d_N} 
||S_j^N||_{h_N}^2) = \frac{\sqrt{-1}}{2\pi N} \partial_b \bar{\partial}_b \log
\Pi_N(z,0,w,0). \end{array} \end{equation}
Together with Proposition \ref{BERGMAN} this completes the proof. \qed

\begin{cor} The statement of Corollary \ref{DEMAILLY} is equivalent to: $
||\bar{\partial}_b \partial_b \log
\Pi_N(x,x)||_{C^k} = O(1).$ \end{cor} 

\section{Parametrix for the Cauchy-Szeg\"o kernel}

Now we recall the necessary background
on the Szeg\"o kernel $\Pi(x,y)$ for a strictly pseudoconvex domain.
The following theorem states that it is a Fourier integral operator of 
complex type, or more precisely a Toeplitz
operator in the sense of Boutet de Monvel-Guillemin \cite{B.G}. The notation below
differs from \cite{B.S} in that $n + 1 = \rm{dim}_{\C} D.$

\begin{theo}\label{BOUTETSJOSTRAND} (\cite{B.S}, Theorem 1.5 and \S 2.c. Let $\Pi(x,y)$ be the Szeg\"o
kernel of the boundary $X$ of a bounded strictly pseudo-convex domain $\Omega$ in a
complex manifold $L$.   Then there exists a symbol $s \in S^{n}(X \times X \times
\Rr^+)$ of the type $$s(x,y,t) \sim \sum_{k=0}^{\infty} t^{n-k} s_k(x,y)$$
so that
$$\Pi (x,y) = \int_0^{\infty} e^{it \psi(x,y)} s(x,y,t) dt$$
where the phase $\psi \in C^{\infty}(D \times D)$ is determined  by
the following properties:

$\bullet$ $\psi(x,x) = \frac{1}{i} \rho(x)$ where $\rho$ is the defining function
of $X.$

$\bullet$ $d_x'' \psi$ and $d_y' \psi$ vanish to infinite order along the
diagonal.

$\bullet$ $\psi(x,y) = -\overline{\psi(y,x)}.$ \end{theo}

More precisely, the phase is determined up  to a function which vanishes to
infinite order at $x = y.$  The integrals are regularized by taking the principal
value (see \cite{B.S}). The second condition 
states that $\psi(x,y)$ is almost analytic. Roughly speaking, $\psi$ is obtained
by Taylor expanding $\rho(z,\bar{z})$ and replacing all the $\bar{z}$'s by
$\bar{w}$'s.  More precisely, the Taylor expansion of $\psi$ near
the diagonal is given by
$$\psi(x + h, x + k) = \frac{1}{i} \sum \frac{\partial^{\alpha + \beta} \rho}
{\partial z^{\alpha}\partial \bar{z}^{\beta}}(x)
\frac{h^{\alpha}}{\alpha!}\frac{\bar{k}^{\beta}}{\beta!}.$$ We note that  Theorem 1.5
of [B.S] is stated only in the case where $L = \C^n$. But in  \cite{B.S}( \S 2.c, 
especially (2.17)) it is extended to general complex manifolds. 

The simplest example is that of the unit ball in $\C^{n + 1}$ in which case the above
formula has the form
$$K_{\partial B}(z,w) = \frac{1}{(1 - \langle z, w \rangle)^{n + 1}} =
\int_0^{\infty} e^{ it \psi_{\partial B}(z,w)} t^{n} dt$$
with $\psi_{\partial B}(z,w)  = 1 - \langle z, w \rangle.$

The above result states that $\Pi$ is a Fourier integral operator with complex
phase, in the class $ I^0_c (X \times X, C^+)$, where $C^+$ is the canonical relation
 $C^+ \subset
T^*X \times T^*X$ generated by  the phase $t \psi(x,y)$
on $X \times X \times \Rr^+.$  Its critical points are the solutions of
$\frac{d}{dt}(t \psi) = 0$, i.e. $\psi = 0$ and on the diagonal of $X \times X$
one has 
\begin{equation} d_x \psi = - d_y \psi = \frac{1}{i} d' \rho|_X. \end{equation}

In particular, the real points of $C^+$ consist in the diagonal of $\Sigma^+ \times
\Sigma^+$ where $\Sigma^+ = \{(x, r \alpha): r > 0\}$ is the cone generated by
the contact form $\alpha = \frac{1}{i} d' \rho.$ In the terminology of [B.G],
$\Pi$ is a Toeplitz structure on the symplectic cone $\Sigma^+.$ 

The principal term $s_0(x,y)$ was also determined in [B.S (4.10)],  using that
$\Pi$ is a projection.  On the diagonal one has
\begin{equation} s_0(x,x) d\mu(x)  = \frac{1}{4\pi^n} (\rm{det}\;\; L_X)
||d\rho|| dx\end{equation}
where $L_X = L_{\rho}|_{T' \oplus T'' \cap TX}$ is the restriction of the Levi
form to the maximal complex subspace of $TX.$

\section{Proof of the Theorem \ref{MAIN}}

The weight space projections $\Pi_N$ are Fourier coefficients of $\Pi$ and hence
may be expressed as:
\begin{equation}\Pi_N(x,y) = \int_0^{\infty} \int_{S^1} e^{- i N \theta} 
e^{it \psi( r_{\theta} x,y)}
s(r_{\theta} x,y,t) dt d\theta \end{equation}
where $r_{\theta}$ denotes the $S^1$ action on $X$. Changing variables $t \mapsto
Nt$ gives
 \begin{equation} \Pi_N(x,y) = N \int_0^{\infty} \int_{S^1}
 e^{ i N ( -\theta +t \psi( r_{\theta} x,y)}
s(r_{\theta} x,y,t N) dt d\theta.\end{equation}

From the fact that Im$\psi(x,y) \geq C (d(x,X) + d(y,X) +
|x-y|^2 + O(|x-y|^3)$ (see \cite{B.S}, Corollary (1.3)) it follows that the phase
\begin{equation} \Psi (t, \theta; x, y) = t \psi(r_{\theta} x,y) - \theta.
\end{equation} has positive imaginary part. Here, $d(x, X)$ is the distance
from $x$ to $X$ and $|x-y|$ is a local Euclidean metric. It follows that
 the integral is a complex
oscillatory integral. Before analysing its asymptotics we simplify the phase.
As above, we choose a
  local holomorphic co-frame $e_L^*$, put $a(z) = |e_L^*|_z^2,$
and write $\nu \in L_z^*$ as $\nu = \lambda e_L^*$. In the associated coordinates 
$(x,y) = (z, \lambda, w, \mu)$ on $X \times X$ we have:
\begin{equation} \rho(z,\lambda) =
a(z)|\lambda|^2,\;\;\;\; \psi(z, \lambda, w, \mu) = \frac{1}{i}
a(z,w) \lambda \bar{\mu} \end{equation}
 where $a(z,w)$ is an almost analytic function on $M \times M$ satisfying
$a(z,z) = a(z)$. On $X$ we  have $a(z) |\lambda|^2 = 1$ so we may write
$\lambda = a(z)^{-\half } e^{i \phi}$. Similarly for $\mu$.  So for $(x,y) =
(z, \phi, w, \phi') \in
X \times X$ we have
\begin{equation} \psi(z, \phi, w, \phi') = \frac{1}{i} (1 - \frac{a(z,w)}{i
\sqrt{a(z)} \sqrt{a(w)}}) e^{i (\phi - \phi')} . \end{equation}
 
\subsection{\bf Proof of Theorem \ref{MAIN}}

 On the diagonal $x = y$ we have $\psi(r_{\theta}x,x) = \frac{1}{i}( 1 -
\frac{a(z,z)}{ a(z)} e^{i \theta})  = \frac{1}{i}(1 - e^{i \theta}).$  So
  \begin{equation} \Psi(t, \theta; x,x) = \frac{t}{i}(1 -   e^{i \theta})  - \theta.
\end{equation} We have
\begin{equation} \begin{array}{l} d_t \Psi = \frac{1}{i} (1 - e^{i \theta} ) \\
d_{\theta} \Psi = t e^{i \theta} - 1 \end{array} \end{equation}
so  
the critical set is ${\cal C} = \{(x, t, \theta): \theta = 0, t = 1\}.$ 
The  Hessian $\Psi''$ on the critical set equals
$$ \left( \begin{array}{ll} 0 & 1  \\
1 & i  \end{array} \right)$$
so the phase is non-degenerate and the Hessian operator is given by 
$L_{\Psi}= \langle (\Psi''(1, 0)^{-1} D,D\rangle =  2  \frac{\partial^2}{\partial t
\partial \theta} - i 
 \frac{\partial^2}{
\partial t^2}.$ It follows by the stationary phase method for complex
oscillatory intgrals that
\begin{equation} \Pi_N(x,x) \sim N \frac{1}{\sqrt{\rm{det} (N \Psi''(1,0)/2 \pi i)}}
\sum_{j,k = 0}^{\infty} N^{n - k - j}  L_j s_k (x,x)\end{equation} 
where $L_j$ is a differential operator of order $2j$ defined by
\begin{equation} L_j s_k(x,x) = \sum_{\nu - \mu = j}\sum_{2 \nu \geq 3 \mu}
\frac{1}{2^{\nu} i^j \mu! \nu!}
 L_{\Psi}^{\nu} [t s_k(r_{\theta} x, x) g^{\mu}(t, \theta)]|_{t= 1,
\theta = 0} \end{equation}
with $g(t,\theta)$  the third order remainder in the Taylor expansion of
$\Psi$ at $(t,\theta) = (1,0).$  More precisely, for any $m \geq 0$, one has by [H I, Theorem 7.7.5] that
\begin{equation}\begin{array}{l} |\Pi_N(x,x) -  N \frac{1}{\sqrt{\rm{det} (N \Psi''(1,0)/2 \pi i)}}
\sum_{j+ k < R}^{\infty} N^{n - k - j}  L_j s_k (x,x) | \\ \\ \leq C N^{n - R}
\sum_{k < R, |\alpha| \leq  2R - 2k} ||D^{\alpha} s_k||_{\infty}.
\end{array} \end{equation}
Note that the hypotheses of [H I, loc.cit.] are
satisfied since the phase has a non-negative imaginary part and since its
critical points are real and  independent of $x$.  Note also that
the expansion can be differentiated any number of times. 
 After some rearrangement, 
the series has the form
\begin{equation} \Pi_N(x,x) =  N^{n} C_n s_0(x,x) + N^{n-1} a_1(x,x) + \dots
\end{equation}
where $C_n$ is a universal constant depending only on $n$ and 
where the  coefficients $s_0(x,x), a_1(x,x) \dots$ depend only on the jets
of the terms $s_k$ along the diagonal.  From the description above of the
leading coefficient $s_0(x,x)$ we have (for some other universal constant $C_n$):
\begin{equation} \Pi_N(x,x)d\mu(x) =  N^{n} C_n \alpha \wedge \omega^n+
O(N^{n-1}).
  \end{equation}
Relative to the Riemannian volume measure $dV_g$, the coefficient is a (non-zero) constant $a_0$ times $ N^{n} $.  Comparing to the leading term of the Riemann-Roch polynomial gives that
$a_0 = 1$,
concluding the proof of (a). \qed

 \subsection{Proof  Corollary \ref{DEMAILLY}}

The new element here is that an arbitrary metric $G$ on $M$, or more precisely its
volume form $dV_G$, is
used to define  orthogonality of sections. We may express $dV_G = J_G \omega^{n}$
for some positive $J_G \in C^{\infty}(M)$ and then express $d\mu_G := d\theta \wedge
\pi^* dV_G = J_G \alpha \wedge d\alpha^n.$  
Let $\Pi^G: L^2(X, d\mu_G) \rightarrow H^2(X, d\mu_G)$ denote the corresponding
orthogonal projection and let $\Pi_N^G$ denote the Fourier components. The
Boutet de Monvel-Sj\"ostrand parametrix construction applies to $\Pi^G$ just as well as to $\Pi$,
the only difference lying in their symbols. The principal symbol for $\Pi^G$ equals
$||d \rho|| det L_X^G$ where  $det L_X^G$ is the determinant relative to $d\mu_G$, 
that is, $det L_X^G = \frac{\omega^n}{d\mu_G} = J_G^{-n}.$ Clearly this is equal
to the determinant of $\omega = Ric(h)$ relative to $dV_G.$ 
 Hence we get
\begin{equation} \Pi_N^G(x,x) = C_n N^n det_G Ric(h) (x)[1 + O(\frac{1}{N})].
\end{equation}

 Corollary \ref{DEMAILLY} follows immediately from this  and from
\begin{equation} |\xi|_{FS, N}^2 = \frac{|\xi|^2}{|S_0^N(x)|^2 + \dots +
|S_N^N(x)|^2} \;\;\;\;\;\;\rm{for}\;\;\;\xi \in E_x^N. \end{equation}
It is equivalent to the statement (cf.[Bou.1, Theoreme Principal] [D, \S 4])
$$\sum |S_j^N(x)|^2   \sim N^n (2\pi)^{-n} |\alpha_1(x) \dots
\alpha_n(x)|$$
where $\alpha_j(z)$ are the eigenvalues of $i c(L) = Ric(h)$ relative to $G$

\subsection{Proof of Corollary \ref{TIAN}} 

Because $\Phi_N$ is a CR map, the asymptotics of the derivatives follow immediately
from the asymptotics of $\Pi_N(x,x)$.  Indeed, $\partial_b \bar{\partial}_b 
\log \Pi_N(x,x) = \partial_b \bar{\partial}_b 
\log \Pi_N(x,y)|_{y=x}.$

By (a)  we have
\begin{equation}\begin{array}{l} \log \Pi_N(x,x) = \log (N^{n} s_0(x,x)
 [ 1 + N^{-1} \frac{s_1}{s_0}
+ \dots])\\ = n \log N + \log s_0(x,x) + \log[1 + N^{-1} \frac{a_1}{s_0}
+ \dots]) = n \log N +  \log s_0(x,x)  + O(\frac{1}{N}). \end{array}
\end{equation} By differentiating the expansion we get
\begin{equation}\partial_b \bar{\partial}_b  \log \Pi_N(x,x) =
 \partial_b \bar{\partial}_b  \log s_0(x,x) + O(\frac{1}{N})= O(1).
\end{equation} \qed

\end{document}